\documentclass[journal]{IEEEtran}
\usepackage{bibentry}  
\usepackage{xcolor,soul,framed,balance, subfigure} 
\pdfoutput=1
\colorlet{shadecolor}{yellow}
\usepackage[pdftex]{graphicx}
\graphicspath{{../pdf/}{../jpeg/}}

\DeclareGraphicsExtensions{.pdf,.jpeg,.png}
\usepackage{dingbat}
\usepackage[cmex10]{amsmath}
\usepackage{array}
\usepackage{mdwmath}
\usepackage{mdwtab}
\usepackage{eqparbox}
\usepackage{url}
\usepackage{siunitx}
\usepackage{booktabs}
\usepackage{pifont}
\usepackage{multirow}
\usepackage{amsmath}
\usepackage{mathtools}
\usepackage{breqn}
\usepackage{graphicx}
\usepackage{array}
\usepackage{setspace} 
\title{\LARGE \bf
Security issues and challenges in V2X: A survey
 }
\author{$^{}$
}

\begin{document}
\bstctlcite{IEEEexample:BSTcontrol}
    \title{Secure OTA Software Updates in Connected Vehicles: A Survey}
\author{Subir Halder, Amrita Ghosal
      and~Mauro Conti,~\IEEEmembership{Senior Member,~IEEE}\\

  \thanks{S. Halder, A. Ghosal and M. Conti are with the Department of Mathematics, University of Padua, via Trieste 63, Padua, 35121, Italy (e-mail: subir.halder@math.unipd.it; amrita.ghosal@math.unipd.it; conti@math.unipd.it).}
 } 
\maketitle

\begin{abstract}
Current trends forecast that Over-the-Air (OTA) software updates will be highly significant for future connected vehicles. The OTA software update will enable upgrading the car functionalities or bug fixations in the embedded software installed on its Electronic Control Units (ECUs) remotely. The introduction of OTA updates in the automotive industry has brought many advantages for both the Original Equipment Manufacturer (OEM) and the driver. According to IHS Automotive, an auto-industry data consulting company, the cost savings from OTA updates for all the OEMs worldwide are estimated to grow to over \$35 billion in 2022. Therefore, many organizations in the automotive industry are presently working on introducing OTA updates as the fundamental feature of their vehicles, considering the growing importance of such kind of updates. However, in terms of security, OTA updates are highly critical as they need complete access to the in-vehicle communication network for installing the latest updates on the ECUs of a vehicle. The main security issues are related to the attacks and threats of wireless connected and update-capable vehicles. 

This survey highlights and discusses remote OTA software updates in the automotive sector, mainly from the security perspective. In particular, the major objective of this survey is to provide a comprehensive and structured outline of various research directions and approaches in OTA update technologies in vehicles. At first, we discuss the connected car technology and then integrate the relationship of remote OTA update features with the connected car. We also present the benefits of remote OTA updates for cars along with relevant statistics. Then, we emphasize on the security challenges and requirements of remote OTA updates along with use cases and standard road safety regulations followed in different countries. We also provide for a classification of the existing works in literature that deal with implementing different secured techniques for remote OTA updates in vehicles. We further provide an analytical discussion on the present scenario of remote OTA updates with respect to care manufacturers. Finally, we identify possible future research directions of remote OTA updates for automobiles, particularly in the area of security.
\end{abstract}

\section{Introduction}
The automobile industry has witnessed a major evolutionary phase, from the generation of manual manoeuvred cars to the ongoing development of connected cars~\cite{Le, Camek, Othmane}. It is apprehended that by 2020, 75\% of the cars in the world will have wireless connectivity~\cite{Khurram16}. The connected car is one of the application's of Internet of Things, that have transformed the driving experience of customers. Connected cars provide high levels of safety and comfort. The car is able to anticipate the current traffic condition due to the enhancement in the degree of automation, that has lead to the reduction in the workload of the driver~\cite{Winter14, Becsi}. The car companies are manufacturing cars having advanced driver assistance systems that can guide the driver's behaviour or provide information about potential crashes so as to reduce the chance of future risks while driving~\cite{Dotzauer15, Dokic15}. The future of connected cars are foreseen as having high automation levels, where all the driving functionalities may be taken over by the car~\cite{Chen16}. The high level of connectivity in connected car introduces a wide range of new security threats as well as privacy concerns~\cite{Checkoway, Constantin, Weise, Henniger}. Therefore, we need a robust security architecture for sustaining the connected car concept. Instances of security compromise by hackers were shown in the past where modern vehicles were attacked using their wireless interfaces~\cite{Foster15, miller}. For example, researchers were able to hack and remotely stop a Jeep Cherokee on a highway~\cite{Miller15} by compromising radio signal, which triggered a recall of 1.4 million vehicles by the Chrysler automobile company. More recently, in another work, researchers were able to compromise and remotely gain control of a Tesla Model S vehicle~\cite{Nie} by compromising wi-fi connectivity, which triggered Tesla to introduce a code signing protection into their cars. These realities of remote cyber-attacks on vehicles has made automobile security as one of the most vital issues~\cite{Liu, Wouters, Roufa}. Security of modern car is a challenging job mainly due to complexity, numerous attack surfaces, and unsafe and old technologies~\cite{Onishi1}. For the sake of convenience, henceforth, we use the term `car', `vehicle' and `automobile' interchangeably.

A modern vehicle contains several Electronic Components Units (ECUs), that perform specific tasks including controlling the functions of engines, controlling the window lifters, windscreen wipers~\cite{Studnia}. The ECUs are interconnected with each other via networks such as, Controller Area Network (CAN)~\cite{King} or Local Interconnect Network (LIN)~\cite{Tuohy}. Since the past decade, the computer based ECUs have substituted many in-vehicle mechanical control systems. In 2013, a study by Frost \& Sullivan found that most cars in those times consisted of 20 to 30 million lines of code for performing certain controlling functions. From the study, Frost \& Sullivan analyzed that with the booming of digitalization and development in technology will increase the investments to \$82.01 billion by 2020~\cite{K079-01}. Software as well as hardware are involved in the development of an ECU and perform the intended functions necessary for a particular module~\cite{Grimm}. Majority of the ECUs in today's market are developed following the V-model. Also, ECU manufacturers are paying attention for developing safety modules on the lines of standards such as, ISO 26262 (functional safety)~\cite{26262} and SAE J3061 (cyber security)~\cite{SAE}. For automotive electric/electronic systems, the ISO 26262 standard adapts from the functional safety standard IEC 61508~\cite{61508}. In addition to ISO 26262 and SAE J3061 standards, two new standards are under development. One of them is ISO 21434-Road Vehicles-Cybersecurity engineering~\cite{Schmittner}. This standard will mainly cover the basic requirements for updating in-vehicle ECUs. Whereas, the second one is being developed by the United Nations Economic Commission for Europe (UNECE) World Forum for Harmonization of Vehicle Regulations (WP.29). Precisely, the Working Party on Automated/Autonomous and Connected Vehicles (GRVA) under WP.29 is developing a recommendation document for cybersecurity and OTA issues~\cite{UNT}.

The Original Equipment Manufacturer (OEM) of the different parts of the vehicles have the obligatory task of providing and managing the software efficiency throughout the lifecycle of the vehicles. The OEM performs the functions of ensuring improved performance as well as rectifying faulty software/firmware (henceforth, software) that may prove detrimental with regard to the safety and security of the vehicles~\cite{Chowdhury}. It is revealed that majority of the recalls are due to software related problems. Therefore, OEM is liable to take extra effort in reducing those problems arising due to software malfunctions. A recent typical example of a recall is that of Volkswagen, where, the company had to recall the 11 millions vehicles it has sold for the last eight years with faulty emissions control software~\cite{Barrett}. Whereas, if the problem was solved using a software update by Volkswagen, the overall process would have involved much less cost, minimized the time taken and also reduced the impact on the environment. Thus, through this real life example we can very clearly understand the significance of remote software updates Over-The-Air (OTA).

Similar to software updates for the smartphones, the software update for various ECUs in connected vehicles is done using OTA technique. It is an efficient and convenient method to update the latest software in the vehicle as this technique saves the visiting time of the customers to repair small bugs in the software. Nevertheless, OTA updates will introduce various new attack vectors~\cite{Checkoway, Larson08, Koscher10}, e.g.,  OBD-II, Bluetooth, Wi-Fi for the malicious actors. Through these attack vectors, malicious attackers exploit the OTA channel to steal the latest updated software, to reprogram ECUs and even control the vehicle remotely. The introduction of new software capabilities has lead to the rapid improvement in crash-avoidance features, such as lane departure warnings and adaptive cruise control. All these new features are modifying the way customers assess cars, giving more importance to user experience and improved safety. These software-reliant systems generally need continuous updates, such as, route map changes, road construction information, changes to safety features. The possibility of sharing unknown situations and solutions collected from other vehicles already on the road, can also enhance safety. OTA updates play a vital role in increasing safety, by keeping driver-assistance features up to date.

\begin{figure}
\centering
  \includegraphics[width=\columnwidth]{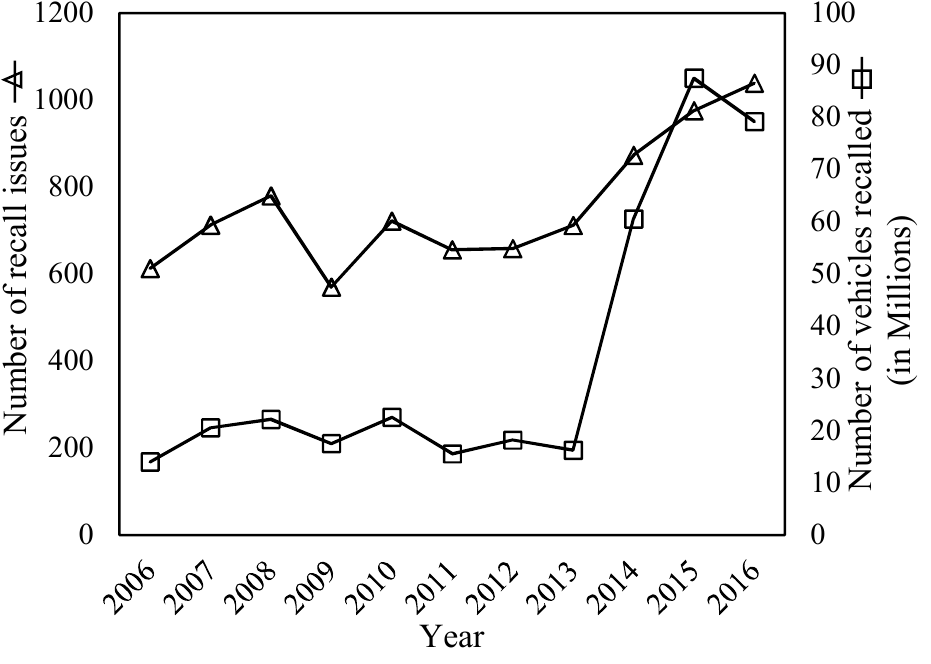}
  \caption{Comparing the number recall issues with number of Vehicles recalled for the year 2006 to 2016.}
  \label{fig:recall}
\end{figure}

Due to the strict vehicle safety standards of various countries~\cite{NHTSA, Europa}, cars are becoming more safe, at the same time more electronic and software dependent than ever before. As the cars are becoming increasingly electronic and software dependent, the number of recalls linked to electronic and software failures has risen exponentially. Figure~\ref{fig:recall} presents the comparison of number of vehicle recall issues occurred due to various faults, e.g., bugs in the software and the number of vehicles affected due to the recall, during the period 2006 to 2016~\cite{Canis}. The plot shows that since 2013, the number of recall issues and number of vehicles recalled has increased significantly. Particularly, the increase in number of recalls is 8.04\%, 22.61\%, 11.79\% and 7\%, in 2013, 2014, 2015 and 2016, respectively. Most importantly, the increase in number of recalled vehicles is 272\%, 44.38\% in 2014 and 2015, respectively. It indicates that although the increase in the number of recall issues is quite less, the number of recalled vehicles increases significantly. Overall, the number of vehicles recalled during 2013 to 2016 ranges between 16.3 to 87.5 millions. Most importantly, another major global marketing information system~\cite{Els}, listed 189 recalls only due to bugs in software during the past 5 years that affected more than 13 million vehicles. Volvo recalled 59,000 cars due to a software issue that caused the engine and the electric system to shut down while the car was in motion~\cite{Volvo}. Honda recalled 350,000 vehicles due to a glitch in the parking brake software~\cite{Honda}. GM recalled 4.3 million cars due to a software issue that blocked the airbags from deploying during an accident~\cite{GM}. All these recalls could have been avoided if there were OTA software updates. OTA updates not only provide assistance for patching against security holes, but also support patching against automotive glitches in software that can cause malfunction in cars. OTA updates provide enormous advantages in keeping in-vehicle software systems up-to-date and maintaining consumer satisfaction. ABI Research, the transformative technology innovation market intelligence institute, predicts almost 203 million OTA supported vehicles will ship by 2022~\cite{ABI}. We list below the various advantages of remote OTA updates in vehicles.

\begin{itemize}
\item \textbf{Lower cost.} OTA updates to systems in near real time, without requiring the owner to bring the car to a dealership or mechanic, helps reduce warranty issues and recalls across potentially millions of vehicles.
    
\item \textbf{Improved safety.} Many under-the-hood systems, such as, steering, braking, and acceleration are electronically actuated. These safety-critical systems can be immediately updated using OTA technology when issues are identified.

\item \textbf{Improved customer satisfaction.} Consumers are spared the inconvenience of bringing their cars to a dealership and they can receive the latest information and safety updates, often without being aware of the change.

\item \textbf{Frequent updates.} Especially, in situations that might otherwise require a recall, OTA updates can be transmitted to all vehicles, whether in the sales lot or on the road. OTA updates allow manufacturers to update software as frequently as necessary in near real time.

\item \textbf{Increased value.} By consistently maintaining in-vehicle software systems with OTA updates, the overall value of the car increases and opens new revenue opportunities to automakers. Also, software configuration costs decreases when multiple software solutions use a single, compatible operating system.
    
\end{itemize}

\begin{figure}
\centering
  \includegraphics[width=\columnwidth]{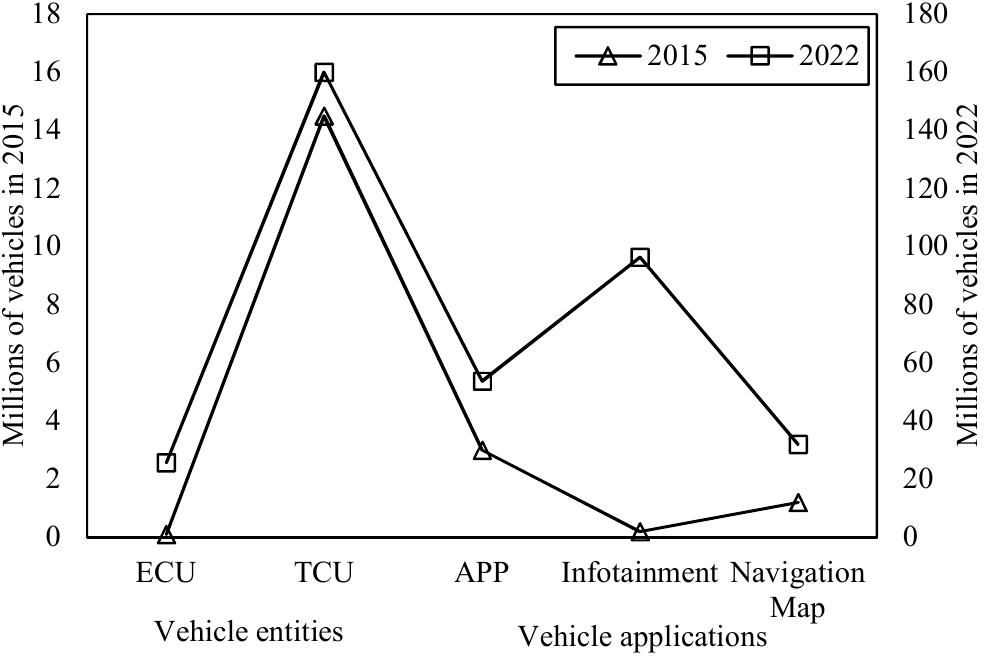}
  \caption{Adaptation of OTA process for updating software in different vehicle entities and application in millions of vehicles during the period 2015-2022.}
  \label{fig:statistic}
\end{figure}

IHS Automotive predicts that automakers will save \$35 billion by using OTA updates in 2022, up from \$2.7 billion in 2015~\cite{IHS}. Figure~\ref{fig:statistic} presents the growth of adoption of OTA for software updation by the automotive industry~\cite{Khurram16}. The automotive industry has chosen five different entities and application, i.e., ECU, Navigation map, Infotainment software application (APP), Infotainment operating software (or, Infotainment) and Telematic Control Unit (TCU) to update their software using OTA process. Interestingly, the table shows that the growth of OTA procedure for software updates in all the five components is exponential.

\paragraph{Contributions} To the best of our knowledge, this survey is the first to demonstrate the significance of OTA updates in vehicular systems. The novelty of this survey is in providing for a detailed description of the current status of OTA updates in vehicles, together with the classification of the existing works and outlining of future research in OTA updates of vehicles. The main contributions of this survey are:
 
\begin{itemize}
\item We discuss the importance of OTA updates in vehicles as well as provide for an overview of OTA updates.
\item We provide a comprehensive discussion on the traditional update and OTA update in vehicles, followed by the issues and challenges faced by OTA updates. In addition, we also discuss the use cases and the standard regulations followed by the countries for vehicle safety. 
\item We classify the existing works in literature, based on their commonality in approaches. 
\item We present a comparative study for the scientific works in literature on vehicle OTA updates, together with the OTA update scenario for car companies.
\item We present promising directions for future research in OTA updates for vehicles, particularly with regard to the security aspects.
\end{itemize}

\paragraph{Organizations} The rest of the paper is organized as follows. In Section II, we provide an overview of OTA update in vehicles, together with discussion on security issues, challenges and requirements. Section II further discusses the use cases and the standard road safety regulations followed by different countries. In Section III, we categorize and discuss the existing security approaches in literature for vehicle OTA updates. Section IV provides a comparative analysis of OTA updates in terms of scientific contributions as well as industrial developments. Finally, we identify the open issues for OTA updates in Section~V and provide conclusion in Section VI.

\section{Overview of OTA Update}
In this section, we present an overview of the OTA update procedure. Specifically, in Section~\ref{Sec:background}, we briefly discuss the software update procedure. Section~\ref{sec:issues} presents the security issues and challenges of the OTA software update technique. We discuss different use cases in Section~\ref{sec:usecases}. Finally, in Section~\ref{sec:regulations}, we describe the standard regulation and type approval regulation compliance laid down by the various countries for safety and security of the passengers.

\subsection{Background}
\label{Sec:background}
\begin{figure}[h]
\centering
  \includegraphics[width=\columnwidth]{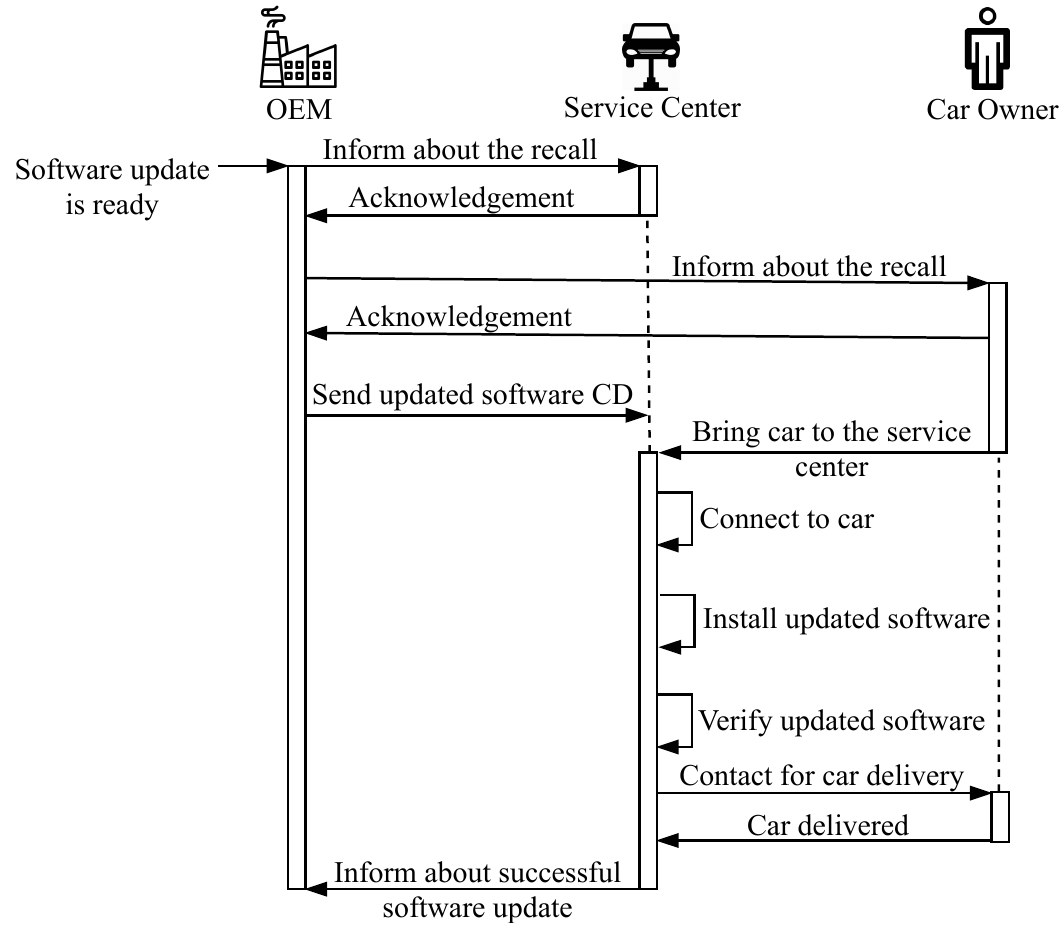}
  \caption{Local software update procedure.}
  \label{fig:localupdate}
\end{figure}

Generally, the entities involved in the ecosystem of software update industry for car are: Car, Cloud Server, Mobile Phone, OEM, Spare part OEM, Software Distributor (SD), Car Owner, Service Center, Insurance Company and Law and Enforcement Personnel~\cite{Cebe}. We broadly categorize the software update procedure into two types: local update and remote update. Here, by local update method, we mean the traditional software update technique, where cars are brought to the service center and a car mechanic updates the software using dedicated tools. Figure~\ref{fig:localupdate} depicts the sequence of steps performed during local software update procedure. The main drawback of local update method is that, it is time and resource consuming, resulting in high cost of labour and customer dissatisfaction~\cite{RS}.

\begin{figure}[t]
\centering
  \includegraphics[width=\columnwidth]{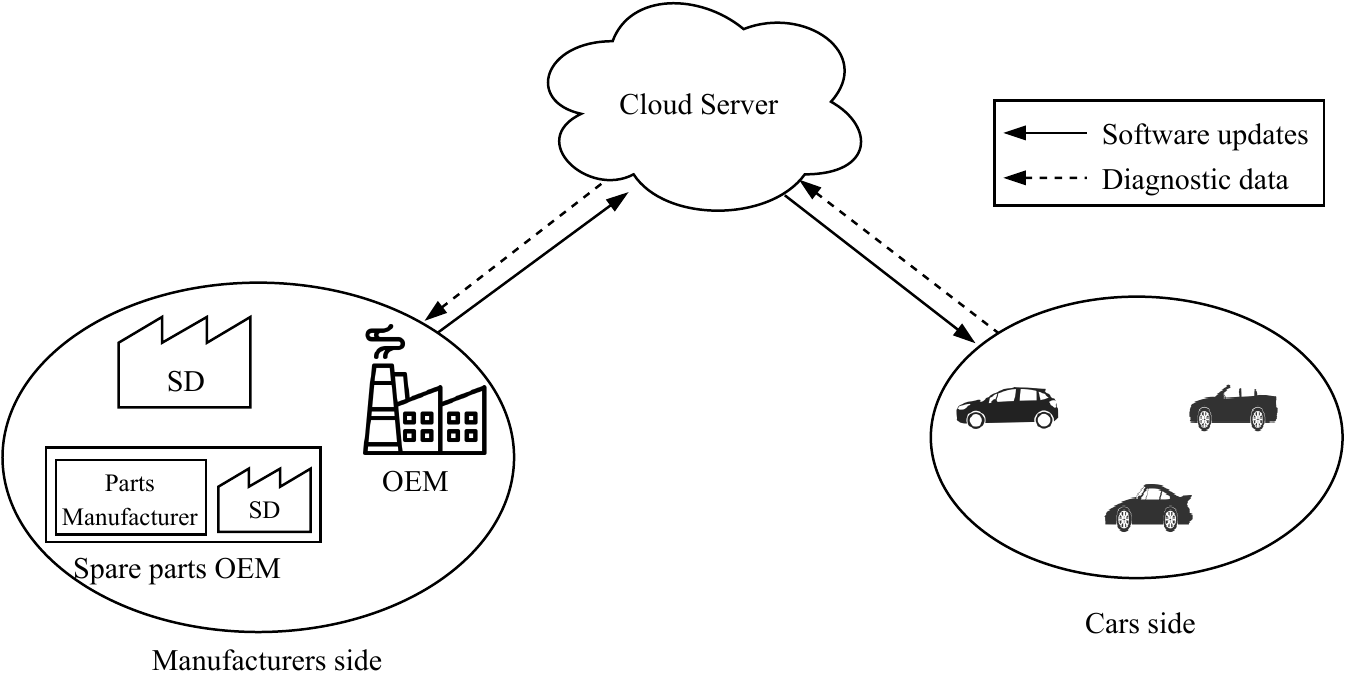}
  \caption{Remote software update procedure.}
  \label{fig:remote}
\end{figure}

On the contrary, in remote update method, connected vehicles allow updating of the software running on their ECUs over-the-air, and the car users can update the software anywhere without having to go to the service center or repair shop. Figure~\ref{fig:remote} illustrates the remote software update procedure. The main advantages of remote update method or OTA-based technique are that, it is fast and cost effective~\cite{Khodari, Odat}. Further, OEM can diagnose the car based on the data received through over-the-air from the car. Remote updates involve complicated techniques that include many steps, such as, secure transfer of the updated transfer along with installation and verification on the specified ECU. Each step of the update process has an impact on the efficiency of the whole update procedure. It is very important for evaluating the effect of various solutions on the vehicle operation and analyzing their performance on real hardware~\cite{Checkoway11}. In future, high complexity of automotive applications will enforce frequent update of electronic devices supporting those applications. Therefore, over-the-air updates are exposed to potential attack surfaces and thus, there is a need for providing stringent requirements with respect to both safety and security in connected vehicular communication~\cite{Kong}.

Recently, with the booming of wireless technology~\cite{Zou, Allam}, secure OTA software updates has been proposed for mobile devices~\cite{Bellissimo, Perito, Kim, Asokan}, specially smartphones~\cite{Barrera, Hotae}. However, there are significant differences between OTA software updates for cars and smartphones in terms of reliability, heterogeneity, usages and installation procedure. The differences are as follows:

\begin{itemize}
\item \textbf{Reliability.} In vehicles, as ECUs are highly interdependent, OTA software updates need more stringent requirements for reliability than OTA software updates for smartphones. If in any case, the vehicle fails to update the software of ECUs properly, the vehicle may not be driven. On the contrary, in smartphones, if any application fails to update properly, then it barely affects the functionality of the other applications. Thus reliability of the OTA software update procedure for vehicles is extremely important compared to the smartphones.

\item \textbf{Heterogeneity.} ECUs in vehicles are highly heterogeneous in terms of memory, computing power and security capability than the smartphones. For example, the ECU for engine control does not have much memory capacity, whereas the ECU for autonomous driving control requires a sufficient memory as it consists of many logic circuits. Therefore, such heterogeneous characteristics of ECUs should be taken into consideration during OTA update procedure in vehicles.

\item \textbf{Installation procedure.} Another major difference is the installation procedure of updated software. Generally, in vehicles, TCU downloads the updated software package from the cloud server. After successful downloading, the TCU at firsts verifies the software package and subsequently distributes the software to the appropriate ECUs. Thereafter, the ECU installs the received software. On the contrary, the smartphone device is in charge of downloading and updating itself. Particularly, application program interface in smartphones, downloads the updated software and installs it into the system.

\item \textbf{Usage.} Diversity in usages of the smartphones and the vehicles play a crucial role during OTA software updation. In the case of a smartphone, it is generally turned-on around the clock and is available for communication throughout the time. Therefore, smartphones can update software any time. On the contrary, some people prefer to drive vehicles on holidays than normal working days. During the normal working days, vehicles may be parked in underground parking areas where communication is quite impossible. If a vehicle takes a long time to switch on, the battery may go dead. Therefore, considering the variety of use situations, it is necessary for the vehicle to shorten the time interval for updating software.

\end{itemize}
    
\subsection{Security Issues, Challenges and Requirements}
\label{sec:issues}
In this section, we discuss the security issues in Section~\ref{sec:II.B.1}. Whereas, Section~\ref{sec:II.B.2} presents the security challenges. Finally, in Section~\ref{sec:II.B.3}, we discuss security requirements necessary for OTA updates in vehicles.

\subsubsection{Security Issues for Vehicular OTA Software Updates}
\label{sec:II.B.1}
For a systematic risk evaluation, it is essential to find out the vital issues of an OTA software update method. For every issue, we identify the relevant aspects that need to be secured. These facts, described below are referred to as security issues.

\begin{itemize}
    \item \textbf{Software update package.} The software package consists of all the required information for updating a particular ECU software. Therefore, the most easy way to attack the ECUs is by tampering with these packages. Specifically, the update package needs to be secured in terms of authenticity and integrity. The freshness of the update information also needs to be preserved by preventing replay attacks.  Also, every update package contains the intellectual property of the system manufacturer. So, the confidentiality of the update package must hold during the update process. The non-repudiation of an update transaction is ensured through tracking of the installation of update packages.
    
    \item \textbf{ECU software.} The ECU software defines the functions of the ECU. The functions of the ECU include the main functionalities along with the applications that support the system’s use cases. Therefore, it is important that the integrity of the software is maintained by ensuring that neither malicious nor unauthorized code gets into the system. The ECU software also needs protection from Denial of Service (DoS) attacks~\cite{Bertolino} for protecting the system availability along with the use cases.
    
    \item \textbf{Vehicle.} The vehicle also indicates the functionality of the system in a more general sense similar to the ECU software. The security properties of integrity and availability is also applicable for vehicles.
    
    \item \textbf{Passenger.} The most important stakeholder is the passenger in environments that interface with human users. With respect to passengers, security related as well as safety critical factors should be taken into consideration. The integrity of the passenger needs to be provided by avoiding injury or physical harm to the passenger.
    
    \item \textbf{Backend.} The backend consists of all the functionalities that are needed for generating the latest software packages. The availability of the update process requires protection for keeping the devices in the field up to date.
    
\end{itemize}

\subsubsection{Challenges for OTA Updates}
\label{sec:II.B.2}
Automakers are looking to deliver increasingly connected systems inside and outside the vehicle. Yet many do not have the technology platform to take advantage of OTA updates, which require a carefully planned methodology to develop, deploy, and maintain. And when a vehicle’s central gateway the entry point into the software systems is receiving and sending information, it is more vulnerable to cyber attacks~\cite{Ashibani}. Therefore, implementing OTA updates in vehicles can present the following challenges.

\begin{itemize}
    \item \textbf{Data integrity.} The update software reaching the vehicles must meet high standards of reliability. The integrity of the update software must be protected so that an adversary is unable to reverse engineer the source code, or steal data of the software. The communication channel over which the software images for updates are received is protected integrity checking, so as to prevent common attacks such as "man-in-the-middle" attacks. Also, the in-vehicle OTA manager which administers the updates, should be protected against manipulation.
    
    \item \textbf{System security.} Cyber threats are dynamic and can rapidly propagate through the in-vehicle network. The OTA update solution must be able to prevent external breaches that could expose vehicle systems and data. The OTA update system should be fully end-to-end secure, otherwise, the impact could be very severe. For example, an automobile company receiving a message that their whole fleet is infected through OTA with ransomware and that the vehicles can be unlocked, only if an instant Bitcoin payment of \$50 million is done.
    
    \item \textbf{Connectivity.} Real-time, nearly continuous OTA updates rely on robust connectivity. As data rates are variable, automakers must ensure that the connection to the vehicle is robust enough to transmit required updates while maintaining low costs. Also. due to the wide differences in bandwidth and signal strengths, the OTA solution must be capable of completing and verifying update downloads that were stopped and restarted several times.
    
    \item \textbf{Standardization.} A lack of industry-wide standards for all OEMs adds to the complexity of software design, interoperability, and connectivity. Handling multiple, independent components and OS configurations requires coordination across vendors and efficient, secure communication. International frameworks that define what the minimum global expectation is for security and for quality of OTA updates, is highly essential.
    
    \item \textbf{Management.} Automobile companies must possess a strong understanding of the software environment to deliver the highly connected experience consumers expect. Today’s cars consist of over one hundred ECUs that need careful design and management of software development cycles. During the vehicle development cycle, software updates need management across a fleet of pre-production vehicles. In addition, the software and systems are usually developed across numerous suppliers and several geographically dispersed design centers. Therefore, the tasks of managing software development cycles, versions, and dependencies become very much significant.
\end{itemize}

\subsubsection{Requirements of OTA Updates}
\label{sec:II.B.3}
In this section, we identify below the security requirements necessary for OTA updates in vehicles.

\begin{itemize}
    \item \textbf{Protection of update package during transmission.} Security must be provided to the update package with respect to authenticity, confidentiality, freshness and integrity while it is being transmitted from the backend to the vehicle.
    
   \item \textbf{Protection of update package while stored.} Before installing the update package received from the manufacturer(s), it must be stored safely. Generally, the target ECU is not utilized during the update process, that might affect the functionality of the vehicle. Thus, an update package received from the manufacturer(s) is kept in a storage within the vehicle until the vehicle is idle for a significantly long duration of time, which is when the latest packages can be safely installed. Therefore, until the duration of time the latest package can be installed, it requires to be stored securely with respect to its authenticity, integrity and freshness, as an adversary may obtain access to the ECU, where the update package is stored and tamper the update package in the storage directly. In addition to authenticity, integrity and freshness, sometimes, the confidentiality of the received update package may need to be preserved in the vehicle.
   
   \item \textbf{Verification of update authorization.} Though the integrity and authenticity of the update package is protected, the instance that issued the update package requires additional verification with regard to possessing of required access rights and authorization of installation of the update at the target location.
   
   \item \textbf{Protection of the update installation.} The delivery and installation of an update package needs to be traceable for many cases. As for example, some update packages may contain errors, resulting in installation problems. In such cases, the owner of the vehicle requires to prove that the error is a result of the erroneous update package installation. Hence, the installation of a genuine update package needs documentation from both the OTA backend and the target ECU.
   
   \textbf{Protection from overloading the backend.} The update packages that are responsible for fixing the safety or security related critical issues in ECUs, should be made easily available to intended vehicles. The OEM should be confident that the distribution backend is safe from security breaches such as, DoS attacks.
   
\end{itemize}

\subsection{Use Cases}
\label{sec:usecases}
In this section, we discuss four use cases which are significant in terms of the OTA updates in vehicles. The four use cases are as follows: recall update process for safety purposes, update operations for non-safety purposes, recall updates for improvements in performance and security risk corrective actions. Recall update process for safety purposes defines the scenarios when a recall take places due to safety related issues and the procedures followed thereafter for informing the concerned authority (e.g., OEM) or the car owner. Update operations for non-safety purposes are those that affect the performance or operation of a vehicle but do not cause any safety related risks to the drive, passengers or pedestrians. Examples of non-recall operation updates include wear and tear of shock absorbers and brake shoes. Performance improvements include the issues that are not related with safety, security or environmental hazards. For example, the use of infotainment app and the navigation maps in cars. Security risk corrective action is concerned with the security risks that the wireless connections are exposed to with regard to the recall process. Example of researchers taking over the control of a Jeep Cherokee remotely~\cite{Miller15}, is very well known. Section~\ref{sec:II.C.1} presents uses case for recall update process. Whereas, in Section~\ref{sec:II.C.2} discusses uses case for non-safety operation updates. We describe the improvement in performance in Section~\ref{sec:II.C.3}. Finally, we present security risk corrective actions in Section~\ref{sec:II.C.4}.

\subsubsection{Recall Update Process for Safety Purposes}
\label{sec:II.C.1}
Most countries have legal requirements that define how the owner of a vehicle is informed about a fault that is related to safety. Every country has a specific definition of a safety defect, though all of them are very much similar. In the United States (US), National Highway Traffic Safety Administration (NHTSA) has a website where recalls are recorded, so that the vehicle owners can have a look and determine whether their vehicle also falls under the recall process. On he other hand, in Europe, a general product safety directive is present, that includes a portion on motor vehicles. However, no provision is provided in the motor vehicles directive for the recall of vehicles that fall under the jurisdiction of an European Union (EU) body. It is the responsibility of the individual member states or countries to check whether the vehicle safety defects are rectified, very similar to the way it is done in the US.  

A defect in a vehicle includes any malfunction in performance, component, material or equipment. A safety defect is generally defined as a problem that is present in the vehicle or in a component of the vehicle that poses a risk on the safety of the vehicle and may be poses security risks for a group of vehicles of similar design, or in list of equipment of the same type and manufacturer. Official recalls include defects such as, malfunctioning of brakes, unexpected braking, unexpected airbag functioning, fuel leak, seatbelt malfunction, detachment of clutch pedal and so on. 

In the US, the responsible authority is the US Department of Transportation National Highway Traffic Safety Administration (US DOT NHTSA)~\cite{Onishi} for delivering vehicle safety standards and notify automobile manufacturers that have safety-related issues or do not satisfy the Federal safety standards. NHTSA also performs the task of monitoring the manufacturer’s remedial action to guarantee that the recall campaign process has been completed successfully. The administrative records for all safety recalls are maintained by the Recall Management Division (RMD). The RMD monitors the recorded recalls for ensuring that the recall is judicious and the accomplishment rate is satisfactory. The NHTSA may raise an investigation of the recall if it suspects the facts indicate a problem related with the recall process execution. A recall process may take place under the following scenarios:

\begin{itemize}
\item Under the initiative of the automobile manufacturer who determines a safety problem;

\item In an investigation based on a court order to recall vehicles, if NHTSA found any safety defect, the automobile manufacturer must follow the below steps:

\begin{itemize}
    \item Report NHTSA about the detail safety defect, including a detail description of what occurs if fault is not attended to and what step is necessary to rectify the fault.
    
    \item Report the vehicle owners through registered mail.
    
    \item Report the dealers and distributors.
\end{itemize}
\end{itemize}

If the vehicle is more than ten years old, the defect must be rectified free of cost. The automobile manufacturer must undertake all means to contact the current owner of the affected vehicles, by utilizing records of its own as well as that obtained from the state vehicle registration records. A website of NHTSA is present, where recalls are listed so that the vehicle owners have the knowledge of the current recalls.

\ The generalized process adopted in case of safety recalls are given below:

\begin{itemize}

\item In case of recalls, either the OEM or the government informs each other whether a recall is needed and is to undertaken.

\item The corresponding ECU supplier is contacted and requested for a new software release. The task of the OEM is to perform tests on the new software for checking the quality of the software.

\item The supplier is responsible for sending the software release to the OEM software update server.

\item The OEM performs the task of identifying vehicles that are affected by the recall. The OEM sends a list containing all the affected vehicles affected due to recall to the OEM Customer Relationship Management (CRM) server. The OEM CRM then associates a vehicle to the dealer who sold the vehicle or to a suitable dealer listed by the customer.
	
\item The OEM CRM sends notification to the dealers for informing them about the required recall along with the list of the recalled vehicles. 

\item The recall update software is sent to all the dealers by the OEM software update server. This is done to facilitate the dealers for preparing their reprogramming tools for updating the software.

\item All customers of recalled vehicles receive a notice from the OEM national level sales company. An update notice is also displayed on the website of the OEM national level sales company.
	
\item The owner of the vehicle brings the vehicle to the dealer shop. In the service bay, the technician connects a serial communication tool to the in-vehicle bus for accessing the targeted ECU. This allows for initiation of the update process of the targeted ECU. The technician performs checks on the targeted ECU for the latest software version for ensuring that proper re-flashing occurred.
	
\item The corresponding customer information is updated in the vehicle and OEM customer database. The update status is reported by the OEM to the government.
\end{itemize}

The following list~\cite{Shavit07} shown below is of car models that experienced software glitches and in most cases were recalled to be re-programmed.

\begin{itemize}

\item 2018: Volvo recalls 16,582 vehicles due to a software glitch in the vehicle connectivity module that cause error in location information to emergency personnel in the event of an accident~\cite{Khaleej}.

\item 2018: Honda recalls 232,000 2018 Accord vehicles and 2019 Insight hybrid cars due to a software glitch that cause malfunctioning of rear camera display~\cite{CNBC}.

\item 2017: FCA recalls 1.25 million trucks due to a software error that might temporarily disable the side air bag and deployment seat of belt pretensioners~\cite{FCA}.

\item 2017: Tesla recalls 53,000 of its Model S and Model X vehicles due to a parking brake issue~\cite{BBC}.

\item 2016: FCA recalls 1.1 million vehicles to add additional transmission control software to prevent inadvertent rollaways. During the process, in a tragic accident, Trek actor Anton Yelchin crushed to death when his Jeep Grand Cherokee rolls down a slope and pins him against a pillar. Yelchin's Jeep was in the recall list but had not been repaired~\cite{Automo}.

\item 2016: Volvo recalls 59,000 cars due to a software fault that cause engine stopping and restarting while the vehicle is in motion~\cite{Volvo}.

\item 2015: Jaguar Land Rover recalls more than 65,000 Range Rover sport utility vehicles due to a software bug that might cause unlatch vehicles’ doors unexpectedly~\cite{BBC1}.

\item 2015: Fiat Chrysler Automobiles recalls 1.4 million vehicles equipped with Uconnect radio head units to fix a software hole that allowed hackers to remotely control various vehicle control systems~\cite{NHTSA1}.

\item 2014: Nissan recalls 990,000 vehicles due to a software problem in the occupant classification system that might cause airbags to not deploy in the event of a crash~\cite{TNYT}.

\item 2014: Honda recalls 175,356 gas-electric hybrid vehicles, in addition to its well-known Fit subcompact, due to a software bug that puts the vehicle at risk of moving or speeding abruptly~\cite{Reuters}.

\end{itemize}

\subsubsection{Update Operations for Non-Safety Purposes}
\label{sec:II.C.2}
There are some problems that has an impact on the performance or operation of the vehicle, but do not affect the safety related to the driver, or passengers of the vehicles or the pedestrians. As for example, problems related to equipments such as batteries, brake pads and shoes, shock absorbers do not effect the safety of the vehicles. No regulations exist in terms of rectification of the problems related with non-safety operations, except those that are covered by the warranty period of the vehicle. 

There is a class of non-safety related issue that comes under the regulations in some countries. Polluting emissions control is one such class of non-safety related issue that falls under this category. In the US, for example, the Environmental Protection Agency (EPA) under the Office of Transportation and Air Quality is accountable for the air pollution compliance check for all motor vehicles. The new cars sold in US must possess an EPA-issued certificate stating that the car conforms to the applicable federal mission standards for controlling air pollution. In Europe, the EU is responsible for defining the acceptable emission standards for exhaust emissions of new vehicles sold in EU member states. The definitions of these standards are present in a series of EU Regulations and Directives. The regulations are applicable for all the member states. Also, the regulations are adopted into the country law as agreed between the Council and the European Parliament. 

Initiation of non-recall updates occur when one or more of the following scenarios arise:

\begin{itemize}
\item The vehicle owner brings the vehicle to the dealer on experiencing some problem. At the dealer's end, it is found that the fault can be rectified using software update. It is observed that the maybe the software update is already available for the application or it is planned for release.
\item The OEM identifies a problem that cannot be experienced by the driver of the vehicle. The ECU supplier requests for the update and it is downloaded to the workstation application. The updated software is installed in the corresponding ECU when the vehicle owner takes the vehicle for regular service; or, 
\item The OEM identifies a problem once it receives a diagnostic trouble code from the vehicle. The dealer contacts the customer and informs him/her that the vehicle can be fixed once it is brought to the dealer or service station.
\end{itemize}

\subsubsection{Performance Improvements}
\label{sec:II.C.3}
Performance improvements consist of all those things that are not related to security, safety, environmental risks. Examples of performance improvements include updating of the infotainment apps by Mercedes Benz and varying the different features of the vehicles by Tesla, such as, rate of acceleration and maximum speed. Another area of performance improvement is navigation map information which is stored on board the vehicle and very quickly gets outdated. If the navigation system is incapable of providing the required route due to outdated data, the reputation of the OEM is at stake. OEMs have pursued vehicle owners with built in navigation systems to pay attention for regular updates, however this attempt did not yield much result. Some OEMs provide for map updates that form part of regular visits to service station, but these updates are restricted to once in a year. Therefore, OTA update of map navigation data is a viable option for vehicle owners, where update take place on a regular basis with no effect on the system performance. BMW is one such company that provides for OTA updates of maps to its customers. Likewise, Tesla can make updates for improvements in acceleration time and location-based air suspension that remembers potholes.

BMW is one such company that at present is providing OTA map updates to its customers. The OTA map update is a standard feature for BMW Connected Drive customers. The Connected Drive backend communicates at regular intervals with the onboard unit of the vehicle for initiating download of map updates. This process allows for requirement of minimal transfer of data. For connectivity, the onboard unit’s internal subscriber identification module is used. The navigation system remains unaffected by the data transfer process. Upon completion of the download, the necessary modifications are applied to the map database.

From the very beginning, Tesla has devised its cars to facilitate powertrain updates for delivery through OTA. This facility was made possible due to the fact that most of the Tesla's vehicles permit ECU to be access using the vehicle’s central telematics system. In July 2015, Tesla announced the Tesla Autopilot feature that allows for supported cars of self steering on roads, change lanes on indication by their user and also find parking spot by themselves. The Tesla Model S vehicles started receiving Autopilot feature from Tesla in the US from September 2015. The OTA software update to Model S Version 7.0 takes advantage of the additional detection features included in Tesla vehicles manufactured since October 2014. The owners of the vehicles were informed that the design of new features was done for enhancing the driver's confidence behind the wheel and minimize the driver's workload as well as avoidance of hazards.

\subsubsection{Security Risk Preventive Measure}
\label{sec:II.C.4}
Previous research have shown that current wireless connections enable hacking of vehicles as well as take control of the vehicle locks and brakes. Recently, researchers~\cite{Miller14} were successful in breaking all the security features of Fiat Chrysler Automobiles and Sprint. In particular, the researches compromised the UConnect onboard systems of Fiat Chrysler vehicles and wireless network, so as to take control over the critical functions of a Jeep Cherokee~\cite{Miller15, Pagliery15}. The attackers at first took control over the radio, the climate controls and the windshield wipers. Next, the attackers moved towards transmission and the brakes. Finally, the car was brought to a completely stop condition in St. Louis, Missouri, US. Andy Greenberg, a journalist of Wired Magazine, who was the driver of the vehicle was a willing victim of this experiment. He shared his experience in Wired Magazine that revealed that he was genuinely frightened when he was helpless in his car and was being remotely controlled by attackers. This demo attack lead to the selection of Jeep Cherokee as the most vulnerable vehicle.

In~\cite{Miller15}, Miller and Valasek found one vulnerable access point that allows anyone having knowledge about the vehicle's IP address obtain access of the chip in the vehicle's head unit. The vehicle’s head unit is the place where the chip’s firmware is rewritten and new code is stored. The new firmware is capable of sending commands through CAN to any of the vital components, like the engine, brakes, sensors or transmission. Prior to the test drive, Miller and Valasek informed Fiat Chrysler Automobile company with sufficient information for enabling the company to notify a recall process for 1.4 million vehicles to repair the security hole in their vehicles. Miller also mentioned that remote updates will provide a new target for hackers. However, he noted that until now, no mischievous hackers have taken over vehicles, and also emphasized on the fact that secured remote updating systems are viable. Though the progress in remotely securing software updates is slow, Miller says, that remote software updates for vehicles are unavoidable in future. With the increase in the number of software in a vehicle, the security risks enhances. Therefore, secure remote updates are becoming the most suitable option.

\subsection{Standard Regulations and Type Approval Regulation Compliance}
\label{sec:regulations}
In this section, we describe the safety regulations followed in different regions of the world. Particularly, Section~\ref{sec:II.D.1} presents the US vehicle safety regulations, while Section~\ref{sec:II.D.2} discusses EU vehicle safety regulations. Finally, Section~\ref{sec:II.D.3} discusses vehicle safety regulations for Asia and Pacific regions.

\subsubsection{US Vehicle Safety Regulations}
\label{sec:II.D.1}
In the very early years of the automobile sector, vehicles in US were lightly regulated by combining both state and private standards. National regulations were not in existence. The vehicle safety legislation was passed in the form of the National Traffic and Motor Vehicle Safety Act in 1966. Every state in US was bounded by the Highway Safety Act of 1966 to generate a highway safety platform on the lines of federal standards for improvements in accident records system, traffic control and driver performance. Starting from January 1967, all vehicles were issued safety standards following the National Traffic and Motor Vehicle Safety Act of 1966. This resulted in formation of an agency named as National Traffic Safety Agency whose task was implementation of the provisions of the new law. In 1970, the National Traffic Safety Agency was given new name as the National Highway Traffic Safety Administration.

Since its inception, NHTSA has issued several safety standards, that include regulations for tires, brakes, seat belts, and airbags. NHTSA does not perform verification in advance about the standards followed by car and parts manufacturers. The law states that the automobile manufacturer or distributor of a motor vehicle has to declare to the motor vehicle distributor during delivery that the motor vehicle conforms with relevant safety standards. In addition, a tag is permanently attached to the vehicle as the proof of certification of the vehicle. It is the responsibility of the manufacturers for examining their vehicles and are accountable for recalls as well as penalties if they do not abide by the Federal Motor Vehicle Safety Standards prescribed by the NHSTA. NHSTA buys sample vehicles from manufacturers and performs tests on the vehicles. If a non-compliance is observed during the testing phase, the NHSTA asks the manufacturer to recall the vehicle model for rectifying the problem.

NHTSA is also responsible for scrutinizing the vehicle faults that affect safety, in addition to enforcing vehicle safety standards. The NHTSA office reviews and investigates complaints of faults raised by vehicle owners, automobile manufacturers and other relevant sources. If NHTSA confirms any safety fault, in general, the automobile manufacturers initiate a recall process. If the automobile manufacturers are unable to go forward with the recall process, NHTSA initiates the recall process by iteself. Next, the automobile manufacturer perform internal investigations that help in identifying a whether a vehicle follows the federal safety standard or not. The automobile manufacturer can issue a recall process by itself for rectifying a safety defect. The law that is setup for the vehicle safety program should ensure that the automobile manufacturer of defective vehicle or component, notifies the concerned owner and rectify the defect without charges. Generally, in most cases, recalls are initiated by automobile manufacturers, that is influenced either by the NHTSA findings or the automobile manufacturers. As an example, in 2016, out of 1039 proposals for recalls, 92 were issued by automobile manufacturers influenced by NHTSA while the remaining 947 were issued based on the findings of the automobile manufacturer.  

\subsubsection{EU Vehicle Safety Regulations}
\label{sec:II.D.2}
Road safety is a pan European issue and it is resolved through a cohesive manner at the EU, national, regional and local level. In EU, road safety polices are usually designed around three main pillars, namely, vehicles, users (including pedestrians, cyclists and drivers), and infrastructure. The coordination of actions and measures adopted by the different authorities in the various domains (e.g., traffic rules enforcement, health care, education, improvement of infrastructure, vehicle type approval and road worthiness inspections) calls for strategic planning. Road safety policy is best defined and implemented under an overarching strategy that addresses all these aspects. Moreover, road safety stakeholders: road user associations, vehicle manufacturers and suppliers, infrastructure managers, fleet operators and other organizations should play an active role in ensuring road safety. The remarkable progress achieved in the past decades is the result of measures taken in these three areas. Today however, as the reduction of road casualties is stagnating, it is even more evident that further progress can only be achieved by continued improvement across the various domains, including that of vehicle safety. For that reason, the present initiative to significantly improve vehicle safety performance has to be viewed in close relation with several other initiatives.

Reflections on whether and how the relevant policy areas should be amended should be seen as part of the preparation of an EU road safety policy framework for the period 2020-2030 (to be proposed as part of the Third Mobility Package in May 2018). Progress in the reduction of road fatalities and serious injuries on EU roads has stalled in recent years, and a revised framework better adapted to this challenge and to the respective modifications in mobility that are outcomes of societal trends (e.g.,an aging society, more cyclists and pedestrians) is needed. This situation leads to the establishment of a dynamic policy adjustment that is capable of addressing the major challenges in an effective mechanism throughout the entire road safety policies. The framework should follow the Safe System approach. This approach is based on the principles that human beings can and will continue to make mistakes and that it is a shared responsibility for actors at all levels to ensure that road crashes do not lead to serious or fatal injuries. In a safe system approach, the safety of all portions of the system, including road use, vehicles, roads and speeds must be upgraded so that in case one side of the systems fails, other sides will still safeguard the person involved. In addition to enhancing vehicle safety features, the foreseen amendment of two directives on road infrastructure safety management and on minimum safety requirements for tunnels, also aim at significantly reducing the number of fatalities and injuries on EU roads. Thus, the named initiatives not only share a common horizon, but they also interlink as the vehicle technology needs to rely on the infrastructure in order to be operational (e.g., visible road markings to support lane keeping assistance technologies). Also, the overall vehicle safety framework should consider the ongoing developments in connected and automated driving, that are consistently increasing. So, the present exercise is closely linked to Commission's strategy on the European Commission's work on motor vehicle safety. It is particularly concerned with the safety of the passengers of the vehicles and the concerned road users. The European Commission brought in a proposal on 17 May 2018, for a Regulation of the European Parliament as well as of the European Council. The European Commission is taking steps towards enhancing the standard vehicle safety equipment. With increasing progress in automotive safety in recent years, new advanced technologies, e.g., emergency lane keeping system, advanced emergency braking system are being made available to customers to prevent accidents. 

In addition to the role of the European Commission in paradigm shifting of vehicle safety issues, it also proposed other passive and active safety developments. All these safety related enhancements will lead to minimization of injuries due to accidents as well as better safeguard the passengers of vehicles, pedestrians and cyclists. Trucks are also being included by providing them with detection systems, and improved side windows and windscreens. This enhancements will result in reduction of blind spots and decrease accidents concerned with pedestrians and bicyclists. Thus, the new proposed safety measures seek to improve road safety for all the concerned stakeholders. Another safety measure adopted is the eCall. The eCall is an automatic emergency phone call system for motor vehicles. The function of the eCall is to reduce the time it takes for emergency services to arrive at the required place. The automobile manufacturers are required to install the eCall technology in vehicle models from 31 march 2018. The eCall technology is envisioned to save many lives and enhance road safety in Europe. The EU law enforces using of safety belts by all vehicle passengers and chile restraint systems for all children.

\subsubsection{Asia and Pacific Region Vehicle Safety Regulations}
\label{sec:II.D.3}
A small proportion of the world's existing number of motor vehicle and total road network belongs to the Asian and Pacific region. It is worth mentioning that inspite of the small figures with respect to the number of motor vehicles, 235,000 road deaths occur annually in Asia and Pacific regions, which is nearly half of the total 500,000 road deaths that happen annually worldwide. The issue of under reporting in this region has made it difficult to understand the number of people injured through road accidents. Reports point out the number of injured due to road accidents in the order of 3 million to 4 million every year. Death due to road accidents account to second largest cause of deaths for the core age groups (5-44 years). Road accidents cause countries to spend between 1\% to 3\% of their annual Gross Domestic Product. Particularly, in East Asia, North-East Asia and South-East Asia, increasing number of road accidents of road users are a matter of great concern.

Road safety was an issue of serious concern for policymakers in the region prior to the launch of the decade of action for road safety and the 2030 Agenda. To address the issue, in November 2006, the 1st Ministerial Conference on Transport adopted the Ministerial Declaration on enhancing road safety in Asia and Pacific region. The Ministerial Declaration included the aim to save 600,000 lives and to avoid a significant number of severe injuries on the roads of Asia and Pacific region between 2007 to 2015. These goals had not yet been met by 2015, the end of the period covered by the Ministerial Declaration. The mandate for global road safety declared the requirement of renewing existing regional road safety missions, targets and indicators. The modified regional road safety goals, targets and indicators for Asia and Pacific 2016-2020 provides important guidance related to policy design and implementation of the safety objectives. The mandate also serves the purpose of providing tools that are used for assessing the outcomes of the implementation of the road safety rules both at the national and regional levels.  

Its overall objective is to achieve 50\% reduction of serious injuries and fatalities arising due to accidents on the roads of Asia and Pacific region from 2011 to 2020. It consists of eight goals, namely, (i) building safer vehicles and familiarizing responsible vehicle advertising, (ii) creating improved the road safety systems for both regions and nation, (iii) creating road safety a higher priority policy, (iv) safer roads for common users, e.g., pedestrians, older people, persons with disabilities, also building safer roads and reducing the seriousness involved in road accidents, (v) building the Asian Highway transport network, (vi) managing and enforcing, (vii) enhancing collaboration and promoting partnerships, (viii) conducting road safety awareness programs for the drivers and public, particularly young people.

The Regional Action Program for Sustainable Transport Connectivity in Asia and the Pacific, Phase I (2017-2021) includes road safety as one of its thematic areas. In the new Regional Action Program, the immediate objective under the road safety theme is to ensure that countries in the area support each other in order to improve road safety as well as to follow the commitments under the Decade of Action for Road Safety and Sustainable Development. Implementation of well coordinated road safety projects involving multiple sectors is necessary. 

\section{Secure OTA Update Techniques in the Scientific Literatures}
\label{sec:3}
\begin{figure*}[h]
\centering
  \includegraphics[width=7in]{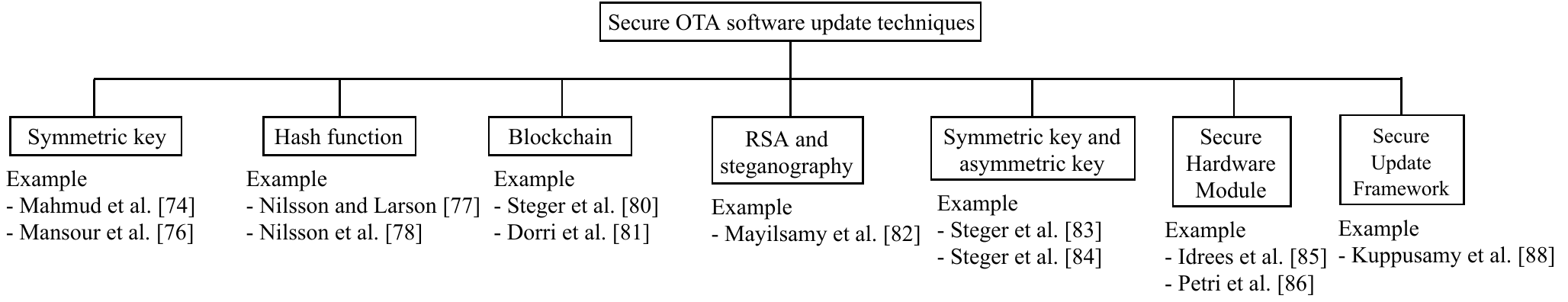}
  \caption{Taxonomy of secure OTA software update techniques.}
  \label{fig:classification}
\end{figure*}

In this section, we discuss about the contributions and achievements of the state-of-the-art secure OTA software update techniques. We classified the existing secure OTA software update solutions into five categories based on the security mechanisms used by the researchers. Figure~\ref{fig:classification} presents the different classifications, which we analyze in this section. Particularly, in Section~\ref{sec:symmetric}, we discuss the symmetric key encryption based OTA software update techniques. Section~\ref{sec:hash} presents the hash function based OTA software update technique. In Section~\ref{sec:blockchain}, we describe the existing blockchain based OTA software update technique. Section~\ref{sec:rsa} discusses the RSA and steganography based secure OTA software update techniques. We discuss the combination of symmetric key and asymmetric key encryption based OTA software update technique in Section~\ref{sec:symasym}. In Section~\ref{sec:HSM}, we put forward the existing Hardware Security Module based OTA software update technique. Finally, we describe a recently published secure software repository framework based OTA update technique in Section~\ref{sec: TUF}.

\subsection{Symmetric Key Encryption Based Techniques}
\label{sec:symmetric}
Mahmud et al.~\cite{Mahmud05} proposed a secure software updation technique for intelligent vehicle. The proposed technique initially shares a set of link keys among the OEM, Software Supplier (SS), and vehicle. Next, before any software updation, a link key is used to establish a secure connection between SS and vehicle for sharing symmetric key. Using the shared symmetric key, the SS encrypts the software and sends it to the vehicle. To increase security, the authors proposed to send the copy of encrypted software at least twice at random intervals. Upon receiving the two copies of the encrypted software, the vehicle first decrypts them and installs one of them. Further, in~\cite{Hossain1}, the authors analyzed the proposed technique to justify the sending of two copies of encrypted software. 

\begin{figure}[h]
\centering
  \includegraphics[width=\columnwidth]{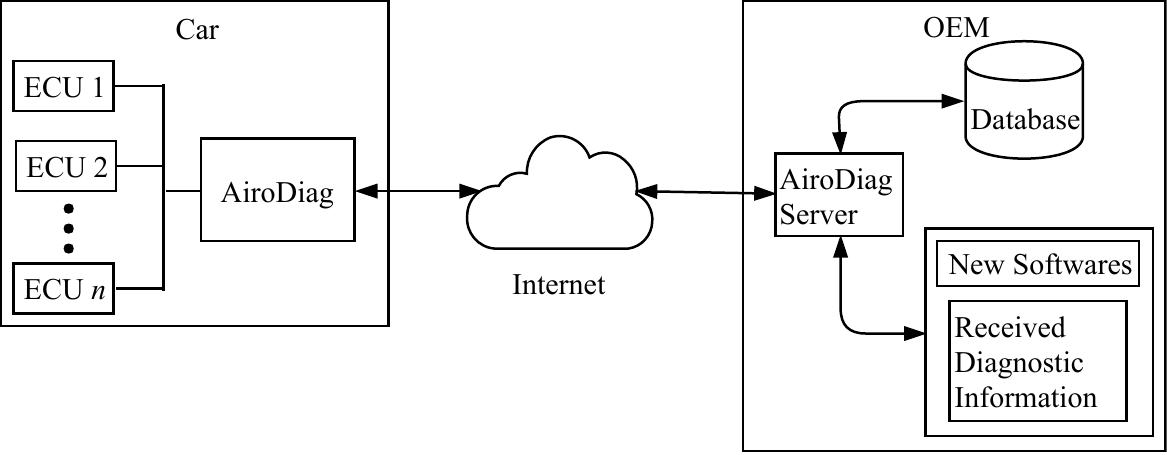}
  \caption{BC-based architecture for secure OTA update.}
  \label{fig:Airo}
\end{figure}

Mansour et al.~\cite{Mansour12} designed a diagnoses and secure software updation system, called AiroDiag for connected vehicle. Figure~\ref{fig:Airo} presents the architecture of the AiroDiag. The main entities in AiroDiag architecture are: OEM, Car and Internet. AiroDiag uses symmetric key technique, particularly, advanced encryption standard to secure the communication during software updation. In AiroDiag, key is stored in the database located in OEM end. The airodiag server always remains connected with the internet, and is kept always ready for any connection request from the car end. In AiroDiag, usually, the software update process is triggered only by the driver/client. Once the software update process is triggered, the vehicle first establishes a secure connection with the OEM. Next, the vehicle informs the OEM about the current version of installed software. In return, OEM sends the new software version for installation. In case, a new software is available, OEM triggers the software update procedure, and establishes a secure connection with the car. The performance of the proposed system is measured through simulation experiments. The results show that based on the size of the software file, the AiroDiag takes 5.14s$\sim$7.85s to update the software in the car.

\subsection{Hash Function Based Techniques}
\label{sec:hash}
\begin{figure}[h]
\centering
  \includegraphics[height=80mm, width=\columnwidth]{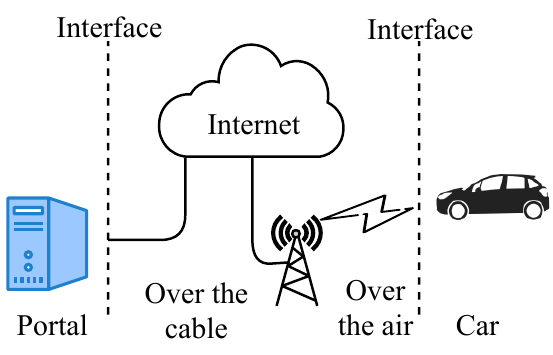}
  \caption{Portal-based communication model for secure OTA update.}
  \label{fig:NilssonICC08}
\end{figure}

Nilsson and Larson~\cite{Nilsson08} proposed a secure OTA firmware update protocol for connected vehicles. The authors considered four entities, i.e., car, portal, Internet and wireless tower in their architecture as shown in Figure~\ref{fig:NilssonICC08}. Here, the portal is the main unit responsible for communication with the cars using wireless connection. In the proposed protocol, first, the updated binary file is divided into number of data fragments. The authors next created a hash chain by hashing each fragment in reverse order. Finally, the portal encrypts each fragment of the hash chain using a pre-shared encryption key before transmitting them to the vehicles. In particular, the portal uses cipher-block chaining as the encryption technique considering the limited resources in the vehicle. Even though the proposed protocol ensures security from eavesdrop, intercept and modification attacks, however, it did not ensure security from denial-of-service attack.

In another work, Nilsson et al.~\cite{Sun08} developed a secure OTA firmware update framework for in-vehicle ECUs. The designed framework comprises of a vehicle and a trusted portal. The trusted portal issues firmware updates including a verification code (generated using a hash function) to the vehicle. The vehicle downloads the firmware using a security protocol. Next, before flashing ECU’s ROM, the vehicle verifies the downloaded firmware using a hardware virtualization technique. Particularly, using hardware virtualization technique, the proposed framework verifies downloaded firmware in two stages, running simultaneously in two systems, namely, control system and functional system. The main strength of the designed framework is that the downloaded firmware is verified before it is flushed into ECU’s ROM. However, the framework fails to detect if the modified firmware is sent from a trusted source.

The authors in~\cite{Roosta08} introduced a key management technique using hash function for software update procedure. The authors used the multicast technique to update the software on a group of nodes. Also, they proposed a rekeying protocol in order to distribute the keys for some specific nodes within the group. However, the proposed technique fails to provide security against number of attacks, including replay attack.

\subsection{Blockchain Based Techniques}
\label{sec:blockchain}
\begin{figure}[h]
\centering
  \includegraphics[width=\columnwidth]{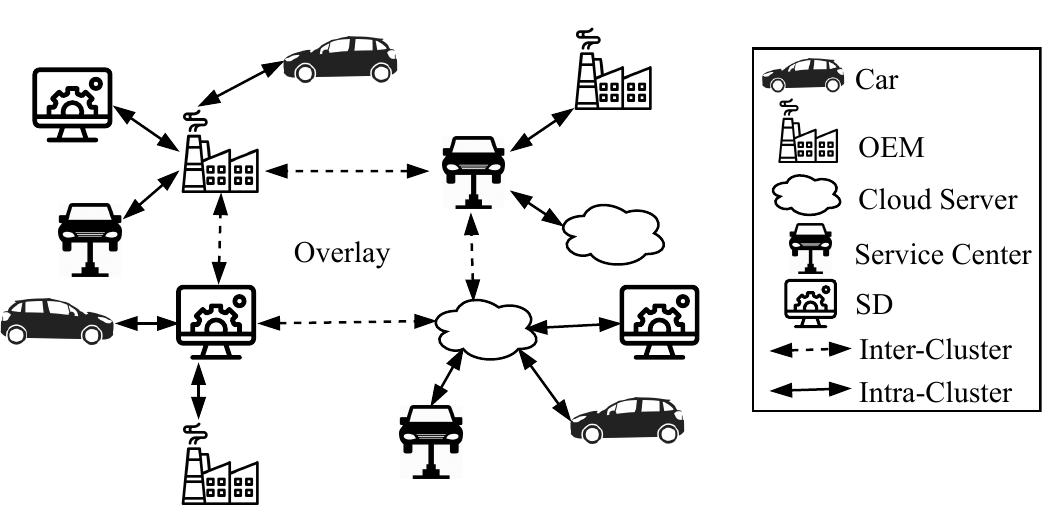}
  \caption{BC-based architecture for secure OTA update.}
  \label{fig:BC}
\end{figure}

Recently, in an interesting work, Steger et al.~\cite{Steger} introduced BlockChain (BC)-based architecture to address the security and privacy issues of OTA software updation for smart vehicles. Figure~\ref{fig:BC} shows the proposed architecture used in this work. The main entities of the architecture are: OEM, service center, car, cloud server and SD. In the proposed architecture, initially, all participating entities form a cluster. Each cluster consists of a number of cluster members and a cluster head. Cluster heads are connected with each other through a overlay network to avoid requirement of any central management. In the proposed architecture, SD triggers the software update process once the new software is developed. First, SD sends a store request with own signature to the cloud server. After successful request validation, cloud server sends an acknowledgement including own signature and a file descriptor required during software uploading process to the SD. Once the new software is uploaded into the cloud server, the SD creates an update transaction in a BC block consisting of information about the location of the new software on the cloud. The SD also includes the public key of the OEM, signs the transaction with private key and ultimately, broadcasts the encrypted transaction to vehicles. To validate applicability of the proposed architecture, the authors presented a proof-of-concept implementation. The results show that the proposed architecture performs better than the certificate-based architecture. Further, in~\cite{Dorri}, Dorri et al. discussed several automotive use cases. They also provided an overview of the probable attack scenarios, and the way the proposed architecture mitigates those attacks.

\subsection{RSA and Steganography}
\label{sec:rsa}
\begin{figure}[h]
\centering
  \includegraphics[height=155mm, width=\columnwidth]{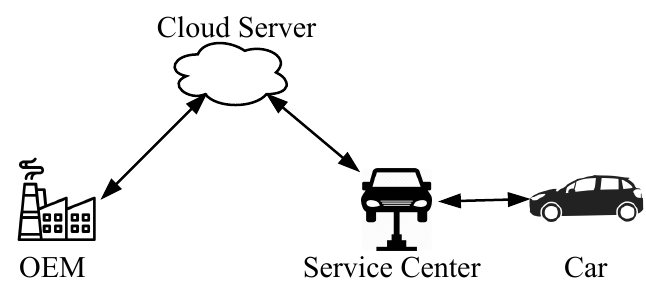}
  \caption{Architecture for secure OTA update used in~\cite{Mayilsamy}.}
  \label{fig:RSAStego}
\end{figure}

Mayilsamy et al.~\cite{Mayilsamy} proposed an integrated method combining cryptography and steganography to secure OTA software update for connected vehicles. Figure~\ref{fig:RSAStego} shows that OEM, service center, car and cloud server are the entities involved in the proposed method. The proposed method uses two levels of security while uploading a new software from OEM to the cloud server. Specifically, the first level, uses a modified RSA algorithm to encrypt the new software file. In the second level, the cipher text of the first level is hidden along the edge regions of the cover image using steganography. Finally, the cloud server stores the stego-image. During the software update process of the car, the service center first downloads the stego-image. Thereafter, the service center performs the necessary decryption and installs the updated software into the car. The authors validate the performance of the proposed technique using MATLAB with Python as backend platform. The simulation results reveal that at the OEM's end, the time required to encrypt file sizes of 1kB, 15kB and 20kB are 3.05s, 6.95s and 8.03s, respectively. On the contrary, at the car's end, the time required to decrypt file sizes of 1kB, 15kB and 20kB are 5.43s, 905.05s and 1590.92s, respectively.

\subsection{Combination of Symmetric Key and Asymmetric Key}
\label{sec:symasym}
\begin{figure}[h]
\centering
  \includegraphics[height=50mm, width=\columnwidth]{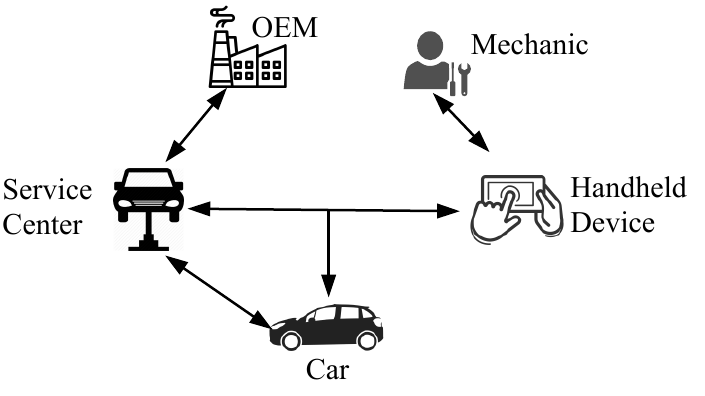}
  \caption{SecUp architecture for secure OTA update.}
  \label{fig:Steger}
\end{figure}

In a state-of-the-art work, Steger et al.~\cite{Steger1} proposed a framework, called SecUp, for secure and efficient OTA software update for connected car. Figure~\ref{fig:Steger} depicts the proposed secure architecture for SecUp. The involved entities of the proposed architecture are: OEM, service center, car and mechanic (with a handheld device). SecUp uses both symmetric and asymmetric key cryptography to secure the OTA software process. Precisely, the authors proposed to use an asymmetric key cryptographic technique, e.g., RSA to secure the communication between service center and car during unicast communication. On the contrary, the authors proposed to use a symmetric key cryptography, e.g., session key based technique for multicast communication from the service center to several cars (through handheld device) in order to enable parallel software update. In SecUp, a mechanic authenticates with the handheld device using a NFC smartcard and a PIN code. Here, service center generates the session key and sends to each car using a unicast data packet encrypted using the public key of car. Next, with the intervention of the mechanic, updated software is transferred to the car from the service center. After receiving the software, the car verifies the software before installation. The authors, in this work, assume that keys are stored in a trusted platform module after a key exchange procedure in a controlled environment and IEEE 802.11s mesh network as communication media. Further, the authors extended the work in~\cite{Steger2} by designing a Wireless Vehicle Interface (WVI) prototype to connect car with the service center. Basically, the function of WVI is to interconnect the vehicular communication bus, e.g., Controller Area Network and ECUs of the connected car with the IEEE 802.11s mesh network. The performance of SecUp is measured through real-world experiments using the Volvo ECU. The results revealed that the update duration for different types of softwares varies between 6.77s$\sim$33.19s. Here, update duration is the time required for transferring the updated software from the service center to the respective ECU.

\subsection{Hardware Security Module}
\label{sec:HSM}
Idrees et al.~\cite{Idrees} proposed a protocol for secure OTA software update for connected car. The proposed protocol exploits the Hardware Security Module (HSM) in order to store the cryptographic key and perform cryptographic operation, e.g., encryption during software update. The network model considered in this work consists of the following entities: OEM, service center and Car. In the proposed protocol, initially, service center remotely diagnoses in-vehicle ECUs. After successful diagnosis, service center sends request to the OEM for decryption key in order to decrypt the downloaded software. OEM sends the key encrypted using the public key of the ECU. The ECU downloads the software from the service center and verifies the authenticity of the downloaded software. If the verification is successful, ECU proceeds with installing the updated software removing the old software. The authors did not provide any qualitative and quantitative analyses of their proposed protocol. Similar to~\cite{Idrees}, in another work, Petri et al.~\cite{Petri} put forward a secure OTA software update mechanism based on Trusted Platform Module (TPM), a kind of HSM. In the proposed technique, initially, the gateway ECU downloads the updated software from the remote server. Next, the ECU validates the downloaded software using pre-defined hash in the TPM. Once the validation is successful, ECU transferred the updated software to the target ECU for installation purpose. The main benefit of using TPM is that it supports computation of many popular cryptographic algorithms, e.g., RSA, SHA, AES~\cite{TCG}. The main limitation of the proposed techniques~\cite{Idrees, Petri} is that, every ECU requires an HSM/TPM, causing additional cost for implementation. 

\subsection{Secure Update Framework}
\label{sec: TUF}

Recently, Kuppusamy et al.~\cite{Karthik} designed a secure OTA software update framework specially for connected cars, called Uptane. In Uptane, an OEM uses a cloud server (i.e., repository) in order to distribute the updated software to ECUs in the form of images and metadata. Here, an image is a collection of code and/or data that allows an ECU to operate properly. Whereas, a metadata contains number of information, e.g., cryptographic hashes and length, to validate the authenticity of an image as well as other metadata. The core of the Uptane design is the adaptation of one of the secure software repository framework The Update Framework (TUF)~\cite{Samuel}. The key principle of TUF is separation of duties in different roles. Specifically, in Uptane, the repository administrator performs different duties via five different roles, i.e., director, root, release, projects and timestamp in order to distribute responsibilities. To increase compromise resilience, repository administrator bundles all updated images and metadata files signed by each of the five roles of the administrator using five private keys and stored in the cloud server. Thereafter, the primary ECU such as, ECU of Telematic Control Unit (TCU) downloads and unpacks the bundle and allocates the updated files to the appropriate secondary ECUs. After receiving the updated files, secondary ECU verifies the image and after successful verification, the old software is updated by the new software. Recently, in two state-of-the-art works~\cite{Asokan} and~\cite{Kuppusamy2}, researchers have pointed out that Uptane is vulnerable to rollback attack due to lack of proper verification mechanism during update software installation at the car end. In case of rollback attack, an ECU installs a different software package version than the most recent version.

\section{Comparative Analysis and Discussion}
In this section, we present a comparative study by analyzing the scientific contributions and industrial developments of OTA updates in cars. Specifically, here we present a comparison of the scientific contributions that deal with secured OTA update techniques in cars. In addition, we also discuss the present scenario of OTA updates related with the different car industries. 

\subsubsection{Scientific Contributions}

We present a comparative study of the various scientific contributions in the field of securing OTA updates in cars through Table~\ref{Table 2}. In Table~\ref{Table 2}, we particularly mention the works that we discussed in Section~\ref{sec:3}. In the scientific literatures, authors used different experimental setups to measure the performance of the proposed techniques, hence posing a difficulty for finding a common platform to evaluate them. Further, due to lack of vivid information, a quantitative analysis of the existing mechanisms is infeasible. Therefore, for shake of convenience, we compare the evaluation setup used in the existing literatures to measure the performance of their proposed mechanism. We compare the evaluation setup based on the three parameters, namely, WVI, ECU and coding technique. It is worth mentioning that WVI and ECU are the hardware component of the connected car. It is worth noting from the Table I that except~\cite{Mahmud05, Nilsson08}, all the other works measured the performance of their proposed techniques through simulation experiment. Further, we notice that except Steger et al.~\cite{Steger1, Steger2}, none of the works used real-world hardware, e.g., WVI and ECU to measure the perform of the proposed mechanism.

\begin{table*}
\caption{Evaluation Setup of Existing Literatures for OTA Updates in Cars}
\label{Table 2}

\begin{tabular}{|p{3.5cm}|p{4cm}|p{4cm}|p{4.5cm}|}
\hline
\textbf{Scientific Literature} & \textbf{WVI} & \textbf{ECU} & \textbf{Coding Technique}  \\ \hline
Mahmud et al.~\cite{Mahmud05, Hossain1}   & Information Not Available  & Information Not Available  & Information Not Available \\ \hline

Mansour et al. \cite{Mansour12}     & Beagle Board (Version xM, Revision B)  & Arduino Duemilanove & Python \\ \hline

Nilsson et al.~\cite{Nilsson08, Sun08}  & Information Not Available      & Information Not Available  & Information Not Available \\ \hline

Steger et al.~\cite{Steger} & BeagleBone Black Board  & Infineon AURIX  & Java and C\\ \hline

Mayilsamy et al. \cite{Mayilsamy} & Information Not Available   & Information Not Available & MATLAB and Python \\ \hline

Steger et al.~\cite{Steger1, Steger2} & BeagleBone Black Board  & Infineon AURIX  & Java \\ \hline

Idrees et al.~\cite{Idrees}   & Information Not Available      & Information Not Available  & Information Not Available \\ \hline

Kuppusamy et al.~\cite{Karthik}   & Information Not Available      & Information Not Available  & Information Not Available \\ \hline
\end{tabular}
\end{table*}

\subsubsection{Industrial Development}
On an average, an automotive vehicle today comprises of approximately 100 ECUs and over 100 million lines of software code. And this number is growing rapidly since the introduction of Connected Cars. Market experts suggest that by 2020, there will be 300 ECUs in a car~\cite{Embitel}, managing most of the functions within a car. In such a complex automotive electronics and software set-up, the need to remotely manage and update the vehicle ECU software becomes all the more important. At a time when the automotive industry is witnessing some disruptive trends, including electrification and autonomous/self-driving vehicles, it is important for OEMs’ to implement efficient software management strategies.

The market reports are clearly suggesting that the future of OTA in automotive looks bright. This is especially applicable for the automotive industry which is shifting its gear towards software driven development of the autonomous systems. The research and market reports have confirmed that OTA updates market is set to grow at a CAGR of 58.15\% during the period 2018-2022. Also, many of the leading automotive players are already on the path of making seamless and secure update of OTA in automotive systems, a reality. In 2017, Bosch had come up with new features required to carry out wireless, OTA updates for cars of the future. These OTA updates ranged from various control units and in-vehicle communication system, to latest encryption techniques and the Bosch IoT cloud.

In this section, we present a comparative study of OTA update characteristics for car companies such as, Tesla, BMW and Mercedes Benz. In addition, we also present a comparative analysis of in-car features that support OTA updates in automobile companies. Particularly, Table~\ref{Table 3} present a comparative analysis of OTA update characteristics for various automobile companies. We notice that only five automobile companies, namely, Tesla, BMW, Mercedes Benz, Audi and General Motors offer OTA software update for their cars. Rest of the companies update the software for their cars locally either in a service center (using dedicated tools) or garage (using USB sticks). Further, we notice that except BMW, all the four automobile companies use 3G data connectivity during OTA software update. However, only Mercedes Benz allows customer/driver to trigger OTA updates.

Table~\ref{Table 4} presents a comparative analysis of in car features, e.g., infotainment, maps and navigation that support OTA updates in automobile companies. We notice that Tesla is the leading automobile company that supports more number of OTA updates for various car features including infotainment, auto emergency break, forward collision warning, power management, maps and navigation. Different from Tesla, other four automobile companies support OTA update for either infotainment or maps and navigation.

\begin{table*}[h]
\caption{Comparative Study of OTA Update Characteristics}
\label{Table 3}
\begin{tabular}{|l|l|l|l|}
\hline
\textbf{Car Companies} & \begin{tabular}[c]{@{}l@{}}\textbf{S/W Update} \\ \textbf{Triggered by Whom}\end{tabular} & \textbf{Update Notification} & \begin{tabular}[c]{@{}l@{}}\textbf{Driving Possibility} \\ \textbf{during Update Process}\end{tabular} \\ \hline

Tesla~\cite{Tesla} & Tesla & \begin{tabular}[c]{@{}l@{}}Sent through an embedded AT\&T 3G data \\ connection or a Wi-Fi router for Model S cars\end{tabular} & No \\ \hline

BMW~\cite{BMW} & BMW & \begin{tabular}[c]{@{}l@{}}Customer receives notification through \\ Connected Drive system present in the car\end{tabular} & No \\ \hline

Mercedes Benz~\cite{MB} & Costumer & \begin{tabular}[c]{@{}l@{}}Update notification sent through an embedded\\ Verizon 3G data connection for C and S class cars\end{tabular} & No \\ \hline

Audi~\cite{Audi} & Information N/A & \begin{tabular}[c]{@{}l@{}}Update notification sent through an embedded \\ T-Mobile 3G data connection for its A3, A4, A5, \\ Q2, Q5 and Q7 cars\end{tabular} & No \\ \hline

General Motors~\cite{GM1} & Information N/A  & \begin{tabular}[c]{@{}l@{}}Chevy Volt model uses the OnStar Verizon 3G \\ data connection for receiving update notification\end{tabular} & No \\ \hline
\end{tabular}
\end{table*}

\begin{table*}[h]
\caption{Comparative Study of the In Car Features that Support OTA Updates}
\label{Table 4}
\scalebox{.87}{
\begin{tabular}{|l|l|l|l|l|l|l|l|l|l|}
\hline
\begin{tabular}[c]{@{}l@{}}\textbf{Car} \\ \textbf{Companies}\end{tabular}  & \begin{tabular}[c]{@{}l@{}}\textbf{Maps and} \\ \textbf{navigation}\end{tabular} & \textbf{Infotainment} & \begin{tabular}[c]{@{}l@{}}\textbf{Power} \\ \textbf{Management}\\ \textbf{Options}\end{tabular} & \begin{tabular}[c]{@{}l@{}}\textbf{Location} \\ \textbf{based} \\ \textbf{Air} \\\textbf{Suspension} \\ \textbf{Settings}\end{tabular} & \begin{tabular}[c]{@{}l@{}}\textbf{Forward} \\ \textbf{Collision} \\ \textbf{Warning}\end{tabular} & \begin{tabular}[c]{@{}l@{}}\textbf{Traffic} \\ \textbf{aware} \\ \textbf{Navigation}\end{tabular} & \begin{tabular}[c]{@{}l@{}}\textbf{Blind} \\ \textbf{Spot} \\ \textbf{Warning}\end{tabular} & \begin{tabular}[c]{@{}l@{}}\textbf{Auto} \\ \textbf{Emergency} \\ \textbf{Braking}\end{tabular} & \begin{tabular}[c]{@{}l@{}}\textbf{Dashcam} \\ \textbf{Feature}\end{tabular} \\ \hline
Tesla       & \checkmark   & \checkmark & \checkmark & \checkmark       & \checkmark & \checkmark & \checkmark & \checkmark & \checkmark                                                           \\ \hline
BMW & \checkmark & \ding{53} & \ding{53} & \ding{53} & \ding{53} & \ding{53} & \ding{53} & \ding{53} & \ding{53}                                                            \\ \hline
\begin{tabular}[c]{@{}l@{}}Mercedes \\ Benz\end{tabular}  & \ding{53}   & \checkmark &  \ding{53} & \ding{53} & \ding{53} & \ding{53} & \ding{53} & \ding{53} & \ding{53}                                                            \\ \hline
Audi & \checkmark & \ding{53} & \ding{53} & \ding{53} & \ding{53} & \ding{53} & \ding{53} & \ding{53} & \ding{53}                                                            \\ \hline
\begin{tabular}[c]{@{}l@{}}General \\ Motors\end{tabular} & \ding{53}   & \checkmark &  \ding{53} & \ding{53} & \ding{53} & \ding{53} & \ding{53} & \ding{53} & \ding{53}                                                            \\ \hline
\end{tabular}}
\end{table*}

\section{Future Scope}
To complete our overview of securing OTA updates for cars, in this section, we present the technical challenges and open directions for future research.

\begin{itemize}
    \item \textbf{Distributed Software Distribution.} The security of the software distribution, from a software provider to an OEM along with the vehicles is still an open area of research. At present, some OEMs, for example, Tesla use Virtual Private Network (VPN) tunnels between the vehicle and the OEM server for OTA updates. This technique supports adequate data protection, but also needs a dedicated communication link between the vehicle and the OEM. The point to point dedicated communication link between the vehicle and the OEM may effect the privacy of the end user. To establish trust within networks, other automotive security architectures make use of certificates. These centralized techniques are not appropriate for highly distributed scenarios comprising of large number of vehicles. So, software distribution during OTA updates need to be more distributed, to safeguard the security primitives required for such a safety critical infrastructure.\\
    
   \item \textbf{Latency Minimization During Software Installation.} Many techniques exist for software installation of the OTA updates, where, a latest image is installed on the ECU. For example, the installation of a latest software image on the ECU through the wired in-vehicle bus network requires more than five times longer than the software distribution using the Block Chain. The latency involved during software installation is of primary concern for such time critical autonomous cars. More research work is needed to be done in this area, that focus om latency minimization during software installation.\\
    
   \item \textbf{Key Management.} In majority of the existing systems, the operational strategies depend on the basic trust involved with pre-programmed keys. Different attacks can be prevented by regular updating of the keys keeping in mind the long lifetime of vehicles. Additionally, the security issues involved in the safe generation and programming of keys are not trivial. Thus, future research work need to focus on the aspect of key management that include operations such as, key generation, key distribution for ensuring secured OTA updates. Also, privacy protection is desirable from the customer point of view. To facilitate privacy protection, a set of anonymous keys could be preloaded to every car. The car owner can utilize a key at a time for a particular time interval, and alter the keys in such a manner that attackers cannot track the key owner. The number of keys that need to be present in the key set, and how frequently the keys should be varied in order to reach a decent level of privacy, needs further thorough investigation.\\
    
\item \textbf{Realization of Parallel or Partial Software Updates.}
Existing works proposed different systems that allowed remote OTA updates by using unicast communications, i.e., a dedicated connection between the OEM and a particular vehicle. Therefore, innovative features like, partial or parallel software updates are not being implemented to a satisfactory level. Contrary to these works, till date, Tesla is the sole vehicle manufacturer offering a solution for OTA software updates for almost all the possible areas. Tesla uses a wireless network to transfer the latest software from the servers located at the OEM side to a Tesla car. However, using the dedicated communication links between the OEM and the car, simultaneous installation of software in several cars and on different ECUs in parallel is not possible.\\

\item \textbf{Ethics and Privacy.} In general, it is expected that the information collected for a particular case is considered according to the ethical information standards based on the specific communication context. A major issue that remains unsolved is how to ensure that the software running in a vehicle is ethical. Therefore, in future, researchers should concentrate on the aspects of ethics and privacy while securing OTA updates for cars.\\

\item \textbf{Vehicle Relationship Management.} An emerging trend in OTA update in cars, is the Vehicle Relationship Management (VRM).
VRM promises to improve vehicle data management through a more transparent and streamlined process. The VRM component consists of VRM Portals (content management, administration and reporting), Cloud-based, multi-tenant VRM platform (with modules for data analytics/management, Vehicle, user and component management) and software agent (resident in target car).
    
\end{itemize}

\section{Conclusion}
A report released by ABI Research says that, by 2022 there would be 22 million vehicles on the road that can get a FOTA upgrade. Thus in the future, firmware and software updates could be more common, more secure and more streamlined.
In this survey paper, we investigate the security issues and challenges of OTA updates in connected vehicles. We provide a comprehensive survey of the underlying concept of OTA updates together with the standard regulations for road safety currently followed by different regions in the world. At first, we provide for a description of the relation of OTA updates with connected cars followed by a brief discussion on the differences between local updates and OTA updates in vehicles. 

Further, we discuss the security issues, challenges and requirements for OTA updates in vehicles. Then, we surveyed the state-of-the-art-works that developed security techniques for carrying out OTA updates in vehicles safely. Next, we present a comparative analysis of the scientific works in literature dealing with vehicle OTA updates. We further provide for a comparative analysis of the current scenario of OTA updates with respect to car companies. Finally, potential research directions for OTA updates in vehicles mainly related to security are identified. This analysis of security issues and challenges of vehicle OTA updates will introduce new and promising perspectives and methodologies for future research in this area.

\section*{Acknowledgment}
This work is partially sponsored by Huawei Innovation Research Program (HIRP), Huawei Technology Co. Ltd., People's Republic of China.

\bibliographystyle{IEEEtran}
\balance
\bibliography{Bibliography}
\end{document}